\documentclass[aps,pre,twocolumn,showpacs,superscriptaddress,groupedaddress,10pt]{revtex4}

\usepackage{graphicx}
\usepackage{color} 
\usepackage{epstopdf}
\usepackage{epsfig}
\usepackage{amssymb}
\usepackage{amsmath}
\usepackage{setspace}
\usepackage{float}

\begin{document}

\title{Formation of dilute adhesion domains driven by weak
  elasticity-mediated interactions}

\author{Nadiv Dharan} \affiliation{Department of Biomedical
  Engineering, Ben Gurion University of the Negev, Be'er Sheva 84105,
  Israel}

\author{Oded Farago} \affiliation{Department of Biomedical
  Engineering, Ben Gurion University of the Negev, Be'er Sheva 84105,
  Israel} \affiliation{Ilse Katz Institute for Nanoscale Science and
  Technology, Ben Gurion University of the Negev, Be'er Sheva 84105,
  Israel}

\begin{abstract}

Cell-cell adhesion is established by specific binding of receptor and
ligand proteins anchored in the cell membranes. The adhesion bonds
attract each other and often aggregate into large clusters that are
central to many biological processes. One possible origin of
attractive interactions between adhesion bonds is the elastic response
of the membranes to their deformation by the bonds. Here, we analyze
these elasticity-mediated interactions using a novel mean-field
approach. Our analysis of systems at different densities of bonds,
$\phi$, reveals that the phase diagram, i.e., the binodal and spinodal
lines, exhibit a nearly-universal behavior when the temperature $T$ is
plotted against the scaled density $x=\phi \xi^2$, where $\xi$ is the
linear size of the membrane's region affected by the presence of a
single isolated bond. The critical point $(\phi_c,T_c)$ is located at
very low densities, and slightly below $T_c$ we identify phase
coexistence between two low-density phases. Dense adhesion domains are
observed only when the height by which the bonds deform the membranes,
$h_0$, is much larger than their thermal roughness, $\Delta$, which
occurs at very low temperatures $T\ll T_c$. We, thus, conclude that
the elasticity-mediated interactions are weak and cannot be regarded
as responsible for the formation of dense adhesion domains. The
weakness of the elasticity-mediated effect and its relevance to dilute
systems only can be attributed to the fact that the membrane's elastic
energy saturates in the semi-dilute regime, when the typical spacing
between the bonds $r\gtrsim \xi$, i.e., for $x\lesssim 1$. Therefore,
at higher densities, only the mixing entropy of the bonds (which
always favors uniform distributions) is thermodynamically relevant. We
discuss the implications of our results to the question of
immunological synapse formation, and demonstrate the
elasticity-mediated interactions may be involved in the aggregation of
these semi-dilute membrane domains.

\end{abstract}

\maketitle

%\vspace{0.45cm}

\section{Introduction}
\label{sec:intro}

The cellular membrane has the ability to adhere to different
biological elements, including the extracellular matrix (ECM), the
cytoskeleton and other cells. Cellular adhesion is mediated by several
adhesion proteins (e.g., cadherins, integrins, and proteins from the
immunoglobulin superfamily) that form specific bonds with receptors
embedded in the adhesive element~\cite{Alberts}. These adhesion bonds
often aggregate into macroscopically large adhesion clusters, such as
focal adhesions, adherens junctions and gap junction
plaques~\cite{Yamada,Harris,Bukauskas}. In addition to providing
mechanical stability to cells, these adhesion domains are essential
for numerous biological processes, including signal
transduction~\cite{Akiyama}, T-cell activation~\cite{Monks}, and
tissue formation~\cite{Vleminckx}. Therefore, it is paramount to gain
a comprehensive understanding of the biophysical principles that
govern the formation of adhesion clusters.

Over the past two decades, many studies have been conducted in order
to better understand the biophysical interactions playing role in the
formation of adhesion clusters~\cite{Sackman,LipowskyRev}. A special
attention has been directed to the effective interactions that are
induced by the membrane elasticity and thermal undulations. These
non-specific interactions have been also studied in relation to
condensation of trans-membrane proteins (membrane
``inclusions'')~\cite{Bruinsma93,Weikl98,Dommersnes}, and in the
broader context of ``Casimir-like'' interactions in condensed
matter~\cite{Kardar99}. Specifically to the problem of adhesion
domains, membrane mediated interactions between adhesion bonds
originate from two interrelated mechanisms operating in concert. The
first mechanism is related to the suppression of membrane thermal
fluctuations by the adhesion bonds, which locally fix the membrane's
height~\cite{FaragoBook}. The resulting loss in the membrane's
fluctuation entropy can be partially mitigated if the adhesion bonds
aggregate into a single domain. The second mechanism stems from local
membrane deformations imposed by the pinning points, which can trigger
a redistribution of the adhesion bonds in order to minimize the
elastic curvature energy~\cite{Rozycki}. Thus, membrane elasticity and
thermal fluctuations induce a potential of mean force (PMF) between
the adhesion bonds. The main challenge in deriving expressions for the
membrane mediated interactions arises from their many-body
character~\cite{Kardar96}, i.e., their non-trivial dependence on the
spatial distribution of the adhesion bonds.

Theoretical studies of membrane mediated interactions are often based
on Helfrich's elasticity theory~\cite{HelfrichA}.  Within the
framework of this model, the membrane is considered as a two
dimensional sheet fluctuating over a flat adhesive surface. Using the
Monge gauge representation and assuming small membrane curvature, the
elastic energy can be expressed by the effective Hamiltonian
\begin{equation}
\label{eq:HH}
{\cal H}=\int\left[\frac{1}{2}\kappa \left(\nabla^2h \right)^2 +
  V(h)\right]d^2{\bf r},
\end{equation}
where $\kappa$ is the membrane's bending modulus, $h=h({\bf r})$ is
the membrane's height (relative to an arbitrary reference plain) at
position ${\bf r}=(x,y)$, and the integration is taken over the
membrane's projected area. The first term in eq.~(\ref{eq:HH}) stands
for the bending energy of the membrane, while the second one denotes a
non-specific confining potential due to interactions between the
membrane and its surroundings, specifically an underlying adhesive
surface. The attachment between the latter and the membrane by $N$
bonds can be incorporated by a set of height constraints satisfying
$h\left(\left\{{\bf r}_i\right\}_{i=1}^N\right)=h_0$, where the bonds
are positioned at $\left\{{\bf r}_i\right\}_{i=1}^N$ and $h_0$ is the
height of the surface. The free energy corresponding to
Hamiltonian~(\ref{eq:HH}) under these constraints constitutes the PMF
between the adhesion bonds.

A commonly used practice in membrane elasticity studies is to assume
that the membrane's free energy has the same form as
eq.~(\ref{eq:HH}), with a renormalized bending modulus and with $V(h)$
representing an effective potential between the surface and the
membrane's mean height profile~\cite{Farago2004}. Using this approach,
Bruinsma, Goulian and Pincus studied the thermodynamics of domains of
gap junctions~\cite{Bruinsma94}. Two regimes with distinct expressions
for $V(h)$ have been proposed, corresponding to different
membrane-surface interactions. In the first regime, coined the {\it
  Helfrich regime}, the bending modulus $\kappa$ is small and,
therefore, thermal fluctuations of the membranes are significant. The
membrane interacts with the surface via thermal collisions, creating
an effective repulsive potential $V(h)\sim
\left(h-h_0\right)^{-2}$~\cite{HelfrichB}. The resulting free energy
has been analyzed within a mean-field picture assuming a lattice of
equally-spaced gap junctions, and was found to grow logarithmically
with the lattice spacing. This result has received support from
coarse-grained membrane simulations~\cite{Farago}. Another prediction
of ref.~\cite{Bruinsma94} was that due to the fluctuation-induced
attraction between the gap junctions, the temperature is renormalized
downward. This prediction was later examined in several computational
studies, which demonstrated that, indeed, the renormalized temperature
is about third to half of the thermodynamic
temperature.~\cite{Weikl,WF,Dharan,Seifert}. These findings highlight
the important role of thermal fluctuations in facilitating conditions
required to adhesion cluster formation. The fact that the renormalized
temperature remains positive implies that in order to achieve
aggregation of adhesion bonds, other attractive interactions must also
be present.
 
The second regime examined in ref.~\cite{Bruinsma94}, termed the {\it
  van der Waals regime}, is characterized by small thermal
fluctuations, which allows one to consider a Lennard-Jones type
potential between the membrane and the surface. For small deviations
from the potential's minimum, a quadratic approximation for
$V(h)=\frac{1}{2}\gamma h^2$ can be assumed. In contrast to the
Helfrich regime where the range of the fluctuation-induced
interactions diverges, the elasticity-mediated interactions in the van
der Waals regime span over a characteristic healing length
\begin{equation}
\label{eq:xi}
\xi=\left(\kappa/\gamma\right)^{1/4},
\end{equation}
beyond which the membrane sets back to the minimum of the confining
potential. As in the Helfrich regime described above, the mean-field
free energy was calculated in ref.~\cite{Bruinsma94} for a lattice
distribution of gap junctions. Coupling the effective interactions
with the adhesion bonds mixing entropy yields the full free energy of
the system, from which conditions for the condensation of adhesion
bonds have been derived.

In the past few years, attempts to develop a more rigorous statistical
mechanical treatment of the van der Waals regime have been
made. Considering the elastic energy given by eq.~(\ref{eq:HH}) with
an harmonic confining potential $V(h)=\frac{1}{2}\gamma h^2$, the
partition function of the system is given by
\begin{equation}
\label{eq:ZN}
Z_N=\int{\cal D}\left[h({\bf r})\right] e^{-\beta{\cal
    H}}\cdot\prod_{i=1}^{N}\delta\left(h({\bf r}_i)-h_0\right),
\end{equation}
where the $N$ pinning points are accounted for through a series of
Dirac-delta functions, and the integration is performed over all
possible height profiles of the membrane. The partition function $Z_N$
can be evaluated by: (i) taking the Fourier representations of the
height function and the Dirac-delta functions, (ii) applying $N$
Hubbard-Stratanovich transformations, and (iii) evaluating the
resulting Gaussian
integrals~\cite{Seifert,Speck2010,Schmidt,Speck2012}. This leads to
the following expression,
\begin{equation}
\label{eq:ZNfinal}
Z_N\simeq\frac{Z_0}{\sqrt{\det
    M}}\exp\left\{-\frac{1}{2}\left(\frac{h_0}{\Delta}\right)^2\sum_{i,j=1}^N\left(M^{-1}\right)_{ij}\right\},
\end{equation}
where $Z_0$ is the partition function corresponding to
Hamiltonian~(\ref{eq:HH}), with $V(h)=\frac{1}{2}\gamma h^2$ and
without ($N=0$) adhesion bonds. The coupling matrix $M$ appearing in
eq.~(\ref{eq:ZNfinal}) is given by,
\begin{eqnarray}
M_{ij}&=&\frac{2k_{\rm B}T}{A_{\rm p}\Delta^2}\sum_{\bf q}\frac{\cos\left[{\bf
      q}\cdot\left({\bf r}_i-{\bf r}_j\right)\right]}{\kappa
  q^4+\gamma}\nonumber \\
 \ &\simeq&-\frac{4}{\pi}{\rm kei}\left(\frac{\left|{\bf r}_i-{\bf
     r}_j\right|}{\xi}\right),
\label{eq:M}
\end{eqnarray}
where the sum runs over all independent Fourier modes ${\bf q}$,
$A_{\rm p}$ is the projected area of the membrane, ${\rm kei(x)}$ is
the Kelvin function~\cite{Abramowitz}, and
\begin{equation}
\label{eq:Delta}
\Delta^2=\langle h({\bf r})^2\rangle=\frac{k_{\rm
    B}T}{8\sqrt{\kappa\gamma}}=\frac{k_{\rm B}T}{8\kappa}\xi^2
\end{equation}
denotes the mean square of height fluctuations (thermal roughness) of
the membrane in the absence of adhesion bonds. For a given
distribution of adhesion bonds, the PMF is given by the free energy
\begin{equation}
\begin{split}
\label{eq:FN1}
&\Phi_N\left(\left\{{\bf r}_i\right\}_{i=1}^N\right) = -k_{\rm
  B}T\ln{\left(\frac{Z_N}{Z_0}\right)}=\\
&\frac{k_{\rm
    B}T}{2}\left[\left(\frac{h_0}{\Delta}\right)^2\sum_{i,j=1}^N\left(M^{-1}\right)_{ij}
  + \ln\left(\det M\right)\right].
\end{split}
\end{equation}
The first term on the r.h.s of eq.~(\ref{eq:FN1}) gives the energy of
the height function that minimizes Hamiltonian~(\ref{eq:HH}) with the
harmonic potential, subject to the height constraints imposed by the
bonds. The second term is the entropic contribution due to the thermal
undulations around this profile~\cite{Schmidt}. Notice that the
energetic and the entropic components in the free energy decouple in
this model, which follows from the quadratic nature of the Hamiltonian
in $q$-space. Also notice that both terms in eq.~(\ref{eq:FN1}) depend
on the elements of the matrix $M_{ij}$~(\ref{eq:M}) in a non-linear
manner, which is a mathematical manifestation of the many body nature
of the PMF.

An interesting observation was made by Speck, Reister and Seifert, who
argued that for small thermal roughness ($\Delta\ll h_0$) the model
depicted by eq.~(\ref{eq:FN1}) belongs to the two dimensional Ising
universality class~\cite{Speck2010}. Furthermore, if the healing
length $\xi$ is smaller than the typical distance between the bonds,
the model can be mapped onto a lattice-gas with nearest neighbor
interactions. By estimating the effective interaction parameter
between adhesion bonds occupying neighboring sites, the authors of
ref.~\cite{Speck2010} were able to draw the phase diagram of the
system and estimate the critical temperature below which clusters
appear.

Despite the insights gained by previous studies, a satisfactory
description of the thermodynamic behavior of the model described by
eq.~(\ref{eq:FN1}) is still lacking. Here, we take another look at
this problem and derive a more accurate picture of the phase diagram,
for a wide range of healing lengths, $\xi$, and adhesion bonds
densities, $\phi$. Our investigation relies on a novel mean-field
treatment of the system's free energy. We obtain the spinodal and
binodal curves and locate the critical temperature of the system,
$T_c$, above which adhesion domains do not form. Results for different
systems exhibit data collapse when $\left(\Delta/h_0\right)^2\sim
T/T_c$ is plotted as a function of the rescaled density
$\xi^2\phi$. Interestingly, we find that the critical point is located
at extremely low densities, which is linked to the many-body membrane
mediated PMF. Therefore, close to critically, a phase coexistence is
found between two extremely dilute phases, while dense domains form
only for $T\ll T_c$, i.e., when each bond deforms the membrane
considerably.

The paper is organized as follows: In section~\ref{sec:MFT} we
introduce our mean-field theoretical treatment. This approach involves
calculations of the elastic energy of systems with randomly
distributed adhesion bonds at various densities. These calculations,
which are described in section~\ref{subsec:Energy}, yield the
expression for the mean-field energy of the system. The free energy is
then obtained by combining the energy with the mean-field mixing
entropy. In section~\ref{subsec:PD}, we analyze the dependence of the
free energy on the density of the bonds, and draw the phase diagram of
the system, i.e., the binodal and spinodal lines. We discuss and
summarize our findings in section~\ref{sec:discussion}.

\section{mean-field theory}
\label{sec:MFT}

The PMF, $\Phi_N$, given by eq.~(\ref{eq:FN1}) corresponds to a system
with a given spatial distribution of $N$ fixed adhesion bonds. The
thermodynamics of a system with $N$ mobile bonds is characterized by
the free energy $F$, which depends on the bond density $\phi=aN/A_{\rm
  p}$, where $a$ is a microscopic unit area for which $0\leq\phi\leq
1$. The free energy $F$ can be derived from the corresponding
partition function $F=-k_{\rm B}T\ln{Z}$, where
\begin{equation}
\label{eq:Zfull}
Z=\underset{\left\{{\bf r}_i\right\}}{{\rm
    Tr}}\left[e^{-\Phi_N\left(\left\{{\bf
    r}_i\right\}_{i=1}^N\right)/k_{\rm B}T}\right],
\end{equation}
is obtained by integrating out the translational degrees of freedom of
the bonds. Since the exact calculation of the partition function is
out of reach, we invoke a simpler mean-field approach. Within a mean
field approximation, the free energy can be written as
%\begin{eqnarray}
\begin{equation}
\begin{split}
\label{eq:MF1}
\frac{aF}{A_{\rm p}}~=~&k_{\rm B}T\left[\phi\ln \phi +
  (1-\phi)\ln(1-\phi)\right]\\
+~&\phi\left\langle\frac{\Phi_N}{N}\right\rangle_{\rm
  MF},
\end{split}
\end{equation}
%\end{eqnarray}
where the first term accounts for the mixing entropy of the bonds, and
the second term represents a mean-field estimation of $\Phi_N$.

We are interested in the so called van der Waals regime (see section
\ref{sec:intro}), which is characterized by small thermal roughness
$\Delta$. Following previous studies~\cite{Bruinsma94,Speck2010}, we
will also make the assumption that each adhesion bond causes a
deformation $h_0$ significantly larger than $\Delta$. This allows us
to drop the second term on the r.h.s.~of eq.~(\ref{eq:FN1}) accounting
for the entropy of the thermal fluctuations, which leaves only the
first term representing the elastic energy of the ground state. The
latter can be estimated by considering a lattice of adhesion bonds
with spacing $r\sim \sqrt{a}\phi^{-0.5}$, which gives an energy
landscape that depends on the ratio $r/\xi$. This approach yields good
analytical expressions for the elastic energy only in the limits
$r/\xi\gg 1$ and $r/\xi\ll 1$ \cite{Bruinsma94}; however, it fails to
capture the correct thermodynamic behavior at the intermediate regime
$r/\xi\sim 1$ where the lattice distribution does not necessarily
represent the energy of a typical random distribution of adhesion
bonds. Here, we take a different approach and derive an empirical
expression for the dependency of the elastic energy on the bonds'
density. We computationally obtain this expression by (i) generating
membranes with random, rather than ordered, distributions of adhesion
bonds, (ii) finding the membrane profile that minimizes the Helfrich
elastic energy of each realization, and (iii) describing the
computational data for the elastic energy by a fitting function, which
applies to the entire range of densities.

\subsection{Energy calculations}
\label{subsec:Energy}

The ground state Helfrich energy corresponding to a random
distribution of adhesion bonds is given by the first term on the
r.h.s.~of eq.~(\ref{eq:FN1}) and can, in principle, be computed by
inverting the coupling matrix (\ref{eq:M}). In practice, this involves
a computationally expensive process and, thus, we adopt a different
strategy based on a direct minimization of the Helfrich
Hamiltonian. This is done by considering a triangular lattice with
lattice spacing $l$. Each site, $i$, represents a small membrane
segment of area $a=\sqrt{3}l^2/2$, and is characterized by a local
height variable $h_i$. On the lattice, $N$ sites are randomly chosen
for the locations of the adhesion bonds, at which we set
$h_i=h_0$. The discrete analogue of the Helfrich
Hamiltonian~(\ref{eq:HH}) is
\begin{eqnarray}
{\cal H_{\rm
    lattice}}&=&\frac{a}{2}\sum_i\left[\kappa\left(\nabla_i^2h_i\right)^2+\gamma
  h_i^2\right]\nonumber
\\ &=&\frac{a\kappa}{2}\sum_i\left[\left(\nabla_i^2h_i\right)^2+
  \left(\frac{h_i}{\xi^2}\right)^2\right],
\label{eq:HHD}
\end{eqnarray}
where the discrete Laplacian at site $i$ is given by
$\nabla_i^2=\left[\frac{2}{3}\sum_{j=1}^6h_j-4h_i\right]/l^2$, with
the sum $j=1\ldots 6$ running over the six nearest neighbors of site
$i$. Starting with $h_i=h_0$ at all sites, we simulate Langevin
dynamics~\cite{GJF} without the noise term (i.e., at zero
temperature), $m\ddot{h}_i=-\alpha \dot{h}_i-\partial{\cal H}/\partial
h_i$, which quickly brings the system to the ground state profile. We
measure all lengths in units of the lattice spacing $l=1$, and the
energy scale is set to $k_{\rm B}T=1$. The density of bonds is given
by $\phi=N/N_s$, where $N_s$ is the number of lattice sites. Most of
the calculations were performed on a triangular lattice of
$104\times120$ sites (with periodic boundary conditions) that has an
aspect ratio close to 1. We calculate the elastic energy of numerous
random realizations at various densities $\phi\leq 0.1$, and for
several values of $\xi$ varying from $\xi=5$ to $\xi=10$. These values
for the correlation length are chosen such that: (i) $\xi$ is
sufficiently larger than the lattice spacing $l=1$, which reduces the
numerical errors associated with the discrete nature of
eq.~(\ref{eq:HHD}) to less than a few percents, and (ii) $\xi$ is much
smaller than the system linear size, to avoid finite size effects.

From eqs.~(\ref{eq:M}), (\ref{eq:Delta}), and (\ref{eq:FN1}) (omitting
the second term on the r.h.s), we infer that for a given set of model
parameters $(\kappa,\ h_0,\ \xi,\ \phi)$, the average elastic energy
has the form
\begin{equation}
\label{eq:fit1}
\left\langle\frac{\Phi_N}{N}\right\rangle_{\rm MF} = \frac{k_{\rm
    B}T}{2}\left(\frac{h_0}{\Delta}\right)^2f(x)
=4\kappa\left(\frac{h_0}{\xi}\right)^2f(x),
\end{equation}
where $f(x)$ is a scaling function of the renormalized density
$x=\xi^2\phi$. Notice that the values of $\kappa$ and $h_0$ can be
fixed arbitrarily since the energy scales like $\kappa h_0^2$ [see
  eq.~(\ref{eq:fit1})], and this scaling behavior is automatically
satisfied by Hamiltonian (\ref{eq:HHD}) which is linear in $\kappa$
and quadratic in $h_i\propto h_0$. The low-density ($x\rightarrow 0$)
asymptotic limit of $f(x)$ is found by considering a system with a
single bond, which gives the energy per bond in dilute systems where
the typical spacing between the bonds is much larger than the
correlation length $\xi$. From eq.~(\ref{eq:FN1}) for $N=1$, we read
that in this limit, $f(x)\rightarrow 1$. In the high density limit,
i.e., when the spacing between bonds is much smaller than $\xi$, the
membrane assumes a nearly flat configuration at height $h_0$. Setting
$h_i=h_0$ in eq.~(\ref{eq:HHD}) and normalizing the energy by the
number of bonds, we obtain the following asymptotic expression
$a\kappa h_0^2/2\phi\xi^4$ for $\Phi_N/N$. Using eq.~(\ref{eq:Delta})
and $a=\sqrt{3}/2$, this yields the decaying form
$f(x)=\sqrt{3}/(16x)$ for $x\gg 1$. Taking these considerations into
account, we propose the following expression for the scaling function
\begin{equation}
\label{eq:f1}
f_1(x)=\frac{1+B_1x}{1+B_2x+\frac{16}{\sqrt{3}}B_1x^2}\,.
\end{equation}
This form ensures the correct asymptotic behavior at low and high
densities, and involves two fitting parameters, $B_1$ and $B_2$, to be
determined by comparison with the numerical data over the entire range
of densities.

\begin{figure}[t]
\begin{center}
{\centering\includegraphics[width=0.5\textwidth]{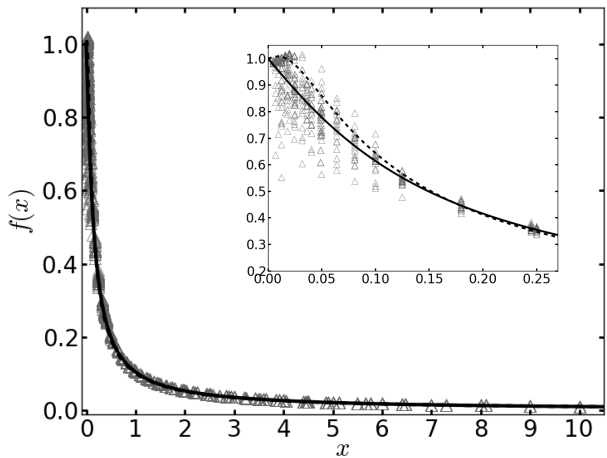}}
\end{center}
\vspace{-0.5cm}
\caption{The scaling function for the elastic energy $f(x)$ [see
    eq.~(\ref{eq:fit1})] as a function of the scaled density $x$. The
  numerical results are presented by triangles. The solid and dashed
  curves depict, respectively, the fitting functions $f_1(x)$
  [eq.~(\ref{eq:f1})] and $f_2(x)$ [eq.~(\ref{eq:f2})] to the
  data. The inset shows an enlarged view of the data and the fitting
  functions for $x\ll1$.}
\label{fig:fig1}
\end{figure}

In Fig.~\ref{fig:fig1} we plot the computational results (triangles)
for the elastic energy per bond, normalized by $4\kappa(h_0/\xi)^2$,
which defines $f(x)$ in eq.~(\ref{eq:fit1}). The data, which is
plotted against the scaled density $x=\xi^2\phi$, exhibits an
excellent data collapse over the entire range $x\leq 10$. The solid
curve represents the fitting of the data to the form $f_1(x)$ given by
eq.~(\ref{eq:f1}), with the parameters $B_1\simeq 5.08$ and $B_2\simeq
9.87 $ that give the best fit. The scatter of the computational data
is due to the randomness of the simulated configurations. As expected,
the scatter is larger for small values $x\ll 1$, where the interaction
between the closer pairs of adhesion bonds dominates the energy of the
configuration. In fact, for some configurations in this regime, we
find $f(x)$ to be slightly larger than unity. This feature is to be
expected, and follows from the non-monotonicity of Kelvin's function
defining the elements of the coupling matrix $M$ [see
  eq.~(\ref{eq:M})]. For $x\ll 1$, the PMF between the bonds can be
approximated by a sum of pair potentials, as was assumed in
ref.~\cite{Speck2012}. By setting $N=2$ in eqs.~(\ref{eq:M})
and~(\ref{eq:FN1}), it is easy to confirm that the pair PMF is
slightly repulsive at large bond separations. We, therefore, conclude
that the scaling function $f(x)$ should be non-monotonic: it first
increases for very small values of $x$, before dropping to zero at
larger values. Furthermore, from the fact that Kelvin's function
converges exponentially to zero for large arguments, one can also
conclude that the derivative of the scaling function $df/dx=0$ at
$x=0$. These features of $f(x)$ in the $x\rightarrow 0$ limit are not
accounted for by the scaling form $f_1(x)$ proposed by
eq.~(\ref{eq:f1}). Therefore, we also consider the three fitting
parameter scaling function
\begin{equation}
\label{eq:f2}
f_2(x)=\frac{1+C_1x+C_2x^2}{1+C_1x+C_3x^2+\frac{16}{\sqrt{3}}C_2x^3},
\end{equation}
which, in contrast to $f_1(x)$, correctly captures the behavior of
$f(x)$ near $x=0$. The fit of the scaling function $f_2(x)$ to the
computational data is also plotted in Fig.~\ref{fig:fig1} (dashed
line) with $C_1\simeq 74.8$, $C_2\simeq 2174$ and $C_3\simeq 1836$
which produce the best fit. The difference between $f_1(x)$ and
$f_2(x)$ is visible only for $x\simeq0$, as seen in the inset in
Fig.~\ref{fig:fig1}. Interestingly, even though $f_2(x)$ is better
suited to represent the scaling function close to the origin than
$f_1(x)$, the latter seems to provide a better fit to the numerical
data. In any case, we expect these two functions to yield similar
binodal and spinodal curves, except for $x\simeq 0$. This will turn
out to be in the vicinity of the critical point, which is where the
validity of the mean-field picture is questionable anyhow.

\subsection{Phase diagram}
\label{subsec:PD}

\begin{figure}[t]
\begin{center}
{\centering\includegraphics[width=0.375\textwidth]{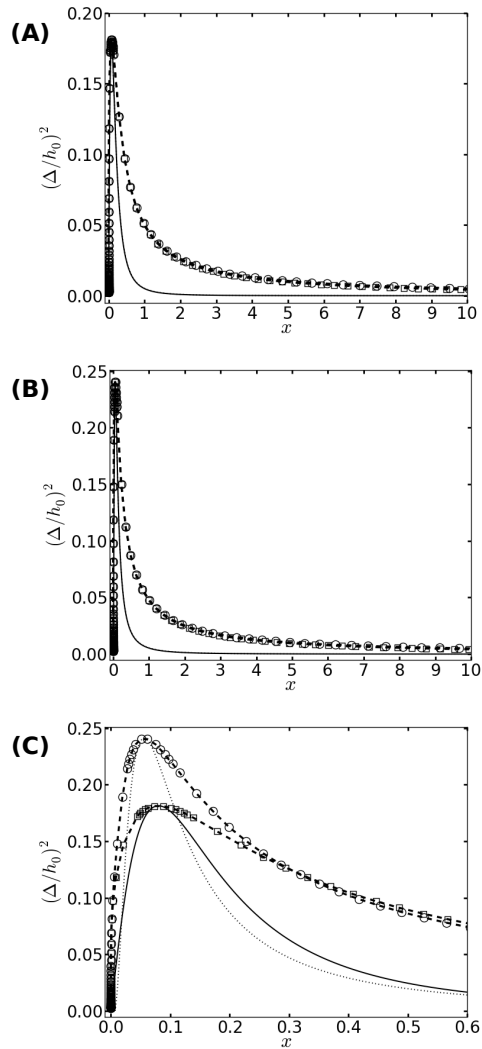}}
\end{center}
\vspace{-0.5cm}
\caption{(A) The phase diagram corresponding to the free energy
  eq.~(\ref{eq:MF3}) with $f(x)=f_1(x)$ given by
  eq.~(\ref{eq:f1}). The binodal curve is represented by the symbols
  (with dashed lines serving as guides to the eye), where squares and
  circles represent data for $\xi=5$ and $\xi=10$, respectively. The
  two binodal curves nearly overlap each other. The spinodal curves,
  which are presented by the solid (for $\xi=5$) and dotted (for
  $\xi=10$) lines, are also indistinguishable. (B) Same as (A), but
  for $f(x)=f_2(x)$ in eq.~(\ref{eq:MF3}). (C) A zoom on the vicinity
  of the critical point, where the differences between the scaling
  functions $f_1(x)$ and $f_2(x)$ are visible. Binodal curves are
  plotted by squares for $f_1(x)$ and circles for $f_2(x)$. The
  spinodal lines are presented by the solid and dotted lines for
  $f_1(x)$ and $f_2(x)$, respectively. The phase diagrams are
  calculated for $\xi=10$.}
\label{fig:fig2}
\end{figure}

Plugging eq.~(\ref{eq:fit1}) into eq.~(\ref{eq:MF1}), the mean-field
free energy, $F$, of a system with adhesion bond concentration $\phi$,
and correlation length $\xi$, reads
\begin{equation}
\label{eq:MF3}
\frac{aF}{A_{\rm p}k_{\rm B}T} \simeq \phi\ln \phi +
(1-\phi)\ln(1-\phi) + 
\frac{1}{2\xi^2}\left(\frac{h_0}{\Delta}\right)^2g(x),
\end{equation}
where $g(x)=xf(x)$. With this expression for $F$, we analytically
obtain the spindoal curve, enclosing the region of thermodynamic
instability, by solving $\partial^2{F}/{\partial\phi^2}=0$, which
yields
\begin{equation}
\label{eq:spin} 
\left(\frac{\Delta}{h_0}\right)^2
=\frac{x\left(x-\xi^2\right)}{2\xi^2} \frac{\partial^2g}{\partial
  x^2}.
\end{equation}
The binodal curve, which defines the thermodynamic coexistence line,
is obtained numerically using a common tangent construction for
$F$. Figs.~\ref{fig:fig2}(A) and (B) show the phase diagrams
calculated using the scaling functions $f_1(x)$ and $f_2(x)$,
respectively. In each of these figures, we plot the spinodal curve for
$\xi=5$ (solid line) and $\xi=10$ (dotted line), which turn out to be
practically indistinguishable. The binodal curves for $\xi=5$ and
$\xi=10$ are given by squares and circles, respectively. As for the
spinodal lines, the binodals for different values of $\xi$ also
overlap each other. Comparing the phase diagrams presented in
Figs.~\ref{fig:fig2}(A) [for $f(x)=f_1(x)$] and (B) [for
  $f(x)=f_2(x)$], we conclude that the phase diagrams appear to be
similar, expect for $x\lesssim0.6$. This is to be expected because
only in this regime, the scaling functions are essentially different
(see inset in Fig.~\ref{fig:fig1}). Fig.~\ref{fig:fig2}(C) presents an
enlargement of the low density regime, showing the binodal [squares
  for $f_1(x)$, and circles for $f_2(x)$] and spinodal [solid line for
  $f_1(x)$, and dotted line for $f_2(x)$] curves, for $\xi=10$. Notice
that the critical point is located at low densities, and the two
scaling functions place it at somewhat different values.

\section{Discussion and Summary}
\label{sec:discussion}

Looking at the phase diagram depicted Fig.~\ref{fig:fig2}, the one
feature that stands out is that the critical point is found at very
low densities. The precise value of the critical scaled density $x_c$
is, of course, unknown since it depends on the form of the scaling
function $f(x)$ [see Fig.~\ref{fig:fig2}(C)], and because the
mean-field picture is not adequate in the vicinity of the critical
point. Nevertheless, it is fair to conclude from the data in
Fig.~\ref{fig:fig2} that $x_c<0.1$, which implies that
$\phi_c=x_c/\xi^2\ll10^{-2}$ (unless the correlation length is
microscopically small, i.e., $\xi\sim 1$). The critical temperature
$T_c$ can be related to the elastic deformation energy due to a single
bond $\Phi_1/k_{\rm B}T=0.5\left(h_0/\Delta\right)^2$. From
Fig.~\ref{fig:fig2} we read that the critical temperature satisfies
$\Phi_1\simeq2-3k_{\rm B}T_c$. Another noticeable feature in
Fig.~\ref{fig:fig2} is the fact that the spinodal and binodal curves
of membranes with different values of $\xi$ overlap each other when
plotted against the scaled density $x$. This does {\em not}\/ a-priori
follow from the data collapse exhibited in Fig.~\ref{fig:fig1},
because of the mixing entropy contribution to the free energy. The
latter depends on the density $\phi$ rather than the scaled density
$x$. At low densities, however, we can use the approximation
$\left(1-\phi\right)\ln\left(1-\phi\right)\simeq -\phi$ in
eq.~(\ref{eq:MF1}), and then it can be easily shown that the spinodal
line [r.h.s of eq.~(\ref{eq:spin})] becomes only a function of
$x$. Thus, the observation in Fig.~\ref{fig:fig2} that the phase
diagram depends on the scaled density is related to the fact that our
investigation focuses on membrane with low densities of bonds.

The fact that the critical point is located at very low densities
means that, slightly below $T_c$, we expect phase coexistence between
two low-density phases. From Fig.~\ref{fig:fig2} we also notice that
for $x\gtrsim1$, phase separation occurs only when the temperature
drops significantly to roughly $T\lesssim 0.2 T_c$. This implies that
low density systems with large $\xi$ will not phase separate unless
the bonds strongly deform the membrane ($h_0\gg\Delta$). In the two
phase region of such a system, the scaled density of the condensed
phase $x\gtrsim 1$ which, depending on the value of $\xi$, could mean
that the density $\phi$ is quite low. We term low-density ($\phi\ll
1$) regions with scaled density $x\sim 1$ as {\em semi-dilute}, and
conclude that the elasticity-mediated interactions may indeed lead to
the formation of such semi-dilute domains.

\begin{figure}[t]
\begin{center}
{\centering\includegraphics[width=0.5\textwidth]{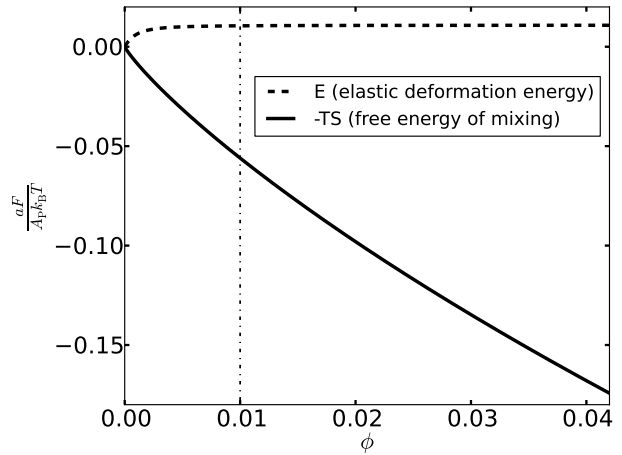}}
\end{center}
\vspace{-0.5cm}
\caption{The free energy, normalized per lattice site and given in
  $k_{\rm B}T$ units, as a function of $\phi$ for $\xi=10$ and
  $(h_0/\Delta)^2=20$. The dashed line is the elastic deformation
  energy, while the solid line represents the free energy of mixing. }
\label{fig:fig3} 
\end{figure}

The ``weakness'' of the elasticity-mediated effect and its inability
to induce formation of dense adhesion domains, can be understood by
looking at the variation of the total elastic deformation energy
[second term on the r.h.s.~of eq.~(\ref{eq:MF3})] with the density of
the bonds $\phi$. The elastic energy $E$, normalized per unit area, is
plotted in Fig.~\ref{fig:fig3} for membranes with $(h_0/\Delta)^2=20$
(corresponding to $T\sim 0.2 T_c$), and $\xi=10$. Also shown in
Fig.~\ref{fig:fig3} is the free energy of mixing $-TS$ ($S$ denotes
the mixing entropy), per unit area, given by the first term on the
r.h.s.~of eq.~(\ref{eq:MF3}). Both contributions to the free energy
are given in units of the thermal energy $k_{\rm B}T$.  We observe
that total elastic deformation energy increases with $\phi$ but,
somewhat surprisingly, saturates at extremely low densities.  The
dashed-dotted vertical line in Fig.~\ref{fig:fig3} at $\phi=0.01$
corresponds to $x=\xi^2\phi=1$, and one can read from the data that
the elastic energy of the membrane barely increases for $x\gtrsim0.5$.
The interpretation of this finding is that one needs a semi-dilute
distribution of about one bond per area $\xi^2$ to cause the membrane
to adopt nearly flat configurations with $h\sim h_0$. Above the scaled
density $x\sim 0.5$, the membrane elastic energy becomes
thermodynamically irrelevant, leaving us with only the mixing entropy
term which always favors uniform distributions. This explains why
phase separation into regions with distinct concentrations of bonds is
possible only at densities below $\phi\sim 0.5\xi^{-2}$. To state the
last conclusion somewhat differently - the elasticity-mediated PMF
induces an attraction between the bonds only if their separation is
{\em larger}\/ than $\xi$. This is an interesting collective
(many-body) effect, exhibiting an ``opposite'' trend compared to the
pair PMF, which is attractive at separations smaller than $\xi$ and is
screened off at larger distances. The pair PMF may play an attractive
role only between two relatively isolated bonds in inhomogeneous
distributions, but such configurations fall outside the framework of
the mean-field picture presented in this work.

To put our findings in a biological context, we look at the example of
the immunological synapse (IS), which forms the contact area between
the T-cell lymphocyte and a target cell. Specifically, the cell-cell
adhesion is mediated via binding between T-cell receptors (TCR) and
MHC-peptide (MHCp) complexes, and between integrin LFA1 and its ligand
ICAM-1~\cite{Grakoui}. These two types of adhesion bonds form a unique
structure, in which TCR-MCHp bonds are clustered in its center, while
the LFA1-ICAM1 bonds aggregate in the periphery of synapse. It is
believed that the central domain, i.e., the TCR-MHCp rich area, plays
a pivotal role in T-cell activation~\cite{Bromley}. Typically, the
bond density within the synapse is around 100 bonds per square
micrometer, and the bond lengths are $14~{\rm nm}$ and $41~{\rm nm}$
for TCR-MHCp and LFA1-ICAM1 bonds, respectively~\cite{Coombs}. We
recall that in the model presented here, $h_0$ represents the local
membrane deformation imposed by a bond relative to the resting height
of the membrane. Thus, if we consider the resting separation between
the two membranes in the IS to be dictated by the longer bonds, we can
estimate the deformation to simply be the difference between the two
bond lengths, $h_0\simeq27~{\rm nm}$. Taking the membrane bending
rigidity to be $\kappa\simeq 15k_{\rm B}T$ and the harmonic potential
strength as $\gamma\simeq 6\cdot10^5k_{\rm B}T~{\rm \mu
  m^{-4}}$~\cite{Chattopadhyay}, we arrive to the values $x\simeq 0.5$
and $\left(\Delta/h_0\right)^2\simeq0.057$ for the coordinates of this
point in the phase diagram displayed in
Fig.~\ref{fig:fig2}. Remarkably, the point lies in the two-phase
region of the phase diagram, close to the binodal line. This raises
the possibility that the TCR-MHCp rich domain may be the semi-dilute
phase coexisting with a dilute phase of vanisingly small
density. Thus, we speculate that the elasticity-mediated interactions
may play an important role in the condensation of the TCR-MHCp
signaling domain. They provide attraction which enables the TCR-MHCp
bonds to spontaneously aggregate into domains with density comparable
to that existing in the IS central zone. This finding is in line with
several recent studies suggesting that passive thermodynamic processes
can describe the short-time condensation of adhesion clusters of the
IS, without evoking any active processes in the cytoskeleton (see,
e.g.,~\cite{mahadevan}, and refs.~therein). Forces stemming from
cytoskeletal activity may be essential during the later stages of IS
pattern formation and
stabilization~\cite{Burroughs,Weikl2009}. Introducing such active
processes into the equilibrium thermodynamic framework presented here
is a task for future studies.

This work was supported by the Israel Science Foundation (ISF) through
grant number 1087/13.

%\newpage

% References

\end{document}